\documentclass[12pt]{iopart}
\usepackage{graphicx,color}      

\begin{document}

\newcommand{\iso}[2]{$^{#1}$#2}
\newcommand{\degC}[1]{$#1^\circ$C}
\newcommand{\Pvap}{P_{\rm vap}}
\renewcommand{\thefootnote}{\arabic{footnote}}

\title[Sub-Doppler modulation spectroscopy of potassium]{Sub-Doppler modulation spectroscopy of potassium for laser stabilization}

\author{L. Mudarikwa, K. Pahwa and J. Goldwin}
\address{Midlands Ultracold Atom Research Centre, School of Physics and Astronomy, University of Birmingham, Edgbaston, Birmingham B15 2TT, UK}
\ead{j.m.goldwin@bham.ac.uk}

\begin{abstract}
We study modulation spectroscopy of the potassium D$_2$ transitions at $766.7$~nm. The vapour pressure, controlled by heating a commercial reference cell, is optimized using conventional saturated absorption spectroscopy. Subsequent heterodyne detection yields sub-Doppler frequency discriminants suitable for stabilizing lasers in experiments with cold atoms.  Comparisons are made between spectra obtained by direct modulation of the probe beam, and those using modulation transfer from the pump via nonlinear mixing.  Finally, suggestions are made for further optimization of the signals.
\end{abstract}

\submitto{\JPB}

\maketitle

\section{\label{sec:Intro}Introduction}

Laser cooling has revolutionized our ability to probe and manipulate atomic gases. In order to perform experiments with cold atoms, lasers first need to be stabilized near strong cooling transitions with residual frequency fluctuations well below the natural linewidth (a few MHz for alkali atoms). The standard technique for obtaining sub-Doppler spectra in room-temperature gases is known as saturated absorption \cite{Han71}. In this method a laser beam is divided into a relatively intense pump and weak probe, which counter-propagate through a gas.  For simplicity, consider the case of two-level atoms.  Since the pump and probe have equal and opposite wave-vectors, atoms with zero Doppler shift along the optical axis will be simultaneously resonant with both beams.  The pump excites these atoms, reducing the ground state atomic population and causing reduced probe absorption in the vicinity of the resonance.  The resulting transmission of the probe therefore shows a Doppler-broadened background with a narrow peak of reduced absorption known as the Lamb dip \cite{Pap80,Ber11}.

For laser stabilization, an electronic frequency discriminant is needed with approximately linear slope around zero volts at the lock point \cite{Fox03,Bec05}.  One can lock directly to the side of a saturated absorption peak by subtracting an electronic offset, but this results in undesirable sensitivity to fluctuations and drifts in optical power, vapour pressure, and offset voltage.  A more stable discriminant can be obtained through phase or frequency modulation and subsequent phase-sensitive detection.  There are a variety of heterodyne techniques which have been used for studying sub-Doppler spectroscopy \cite{Lev79,Bjo80,Raj80,Sny80,Hal81a,Cam82}.  These vary in the details of the optical configuration, the number of frequency components and how they are produced, and the relative strengths and detunings of the contributing beams.  Beams may be generated by phase, frequency, or amplitude modulation, frequency shifting, or from independent sources.  For small modulation depths, and modulation frequencies much smaller than the natural linewidth, phase modulation of the probe results in a signal which is approximately proportional to the derivative of the absorption feature \cite{Bjo83}.  For large modulation frequencies, spectra exhibit multiplet structures, similar to those obtained with the popular Pound-Drever-Hall method for stabilizing lasers to optical cavities \cite{Dre83,Bla01,Fox03}.  

Here we investigate two methods which utilize similar experimental arrangements, but which give rise to very different locking signals.  The first method involves weak phase modulation of the probe beam, with a modulation frequency on the order of the natural linewidth.  This method is similar to traditional lock-in detection \cite{Moo09}; we refer to this as \textit{direct} modulation.  In the second method, only the pump beam phase is modulated.  Through various nonlinear mixing processes, this can generate probe sidebands which beat with the carrier at the detector \cite{Shi82,Blo83,Blo83err}.  This is known as modulation \textit{transfer} spectroscopy.

\begin{figure}
	\centering
		\includegraphics[scale=0.5]{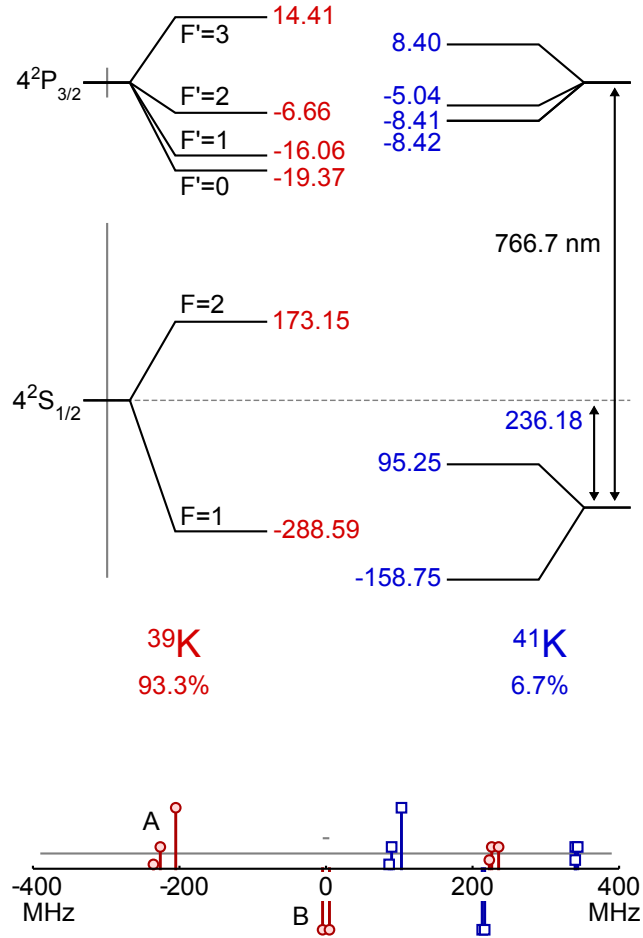}
	\caption{(Colour online.) Energy level schematics for naturally abundant potassium isotopes.  Above: energy level diagrams for \iso{39}{K} and \iso{41}{K}.  Hyperfine frequencies obtained from \cite{Fal06} are given in MHz, with excited states expanded by 10$\times$ for clarity; $F$ is the total (electronic plus nuclear) angular momentum.  The vertical grey bars in the excited and ground states show the natural width $\pm \gamma/(2\pi)$ and the Doppler width $\pm\Delta/(2\pi)$, respectively.  Below: transition frequencies for \iso{39}{K} (red circles) and \iso{41}{K} (blue squares). Upward-pointing features are normal transitions, and downward-pointing features are ground-state crossovers; excited-state crossovers have been omitted for clarity. The heights of normal transitions reflect the relative oscillator strengths. The centre frequency is taken to be the midpoint between the \iso{39}{K} ground state crossovers.  The horizontal grey bars are the natural and Doppler widths as above.}
	\label{fig:schematic}
\end{figure}

Figure \ref{fig:schematic} shows the D$_2$ energy levels for the two most abundant isotopes of potassium, \iso{39}{K} and \iso{41}{K}, with natural abundances of 93.3\% and 6.7\%, respectively.  The natural half-width at half-maximum (HWHM) is \mbox{$\gamma=2\pi\!\times3.017$}~MHz~\cite{Wan97}, which is on the order of the excited-state hyperfine splittings. Additionally, there exist so-called crossover resonances in saturated absorption at frequencies half-way between each pair of excited state transitions.  These features are due to moving atoms whose Doppler shifts equal half the frequency difference between transitions.  This leads to a group of six overlapping features for each hyperfine ground state of each isotope.  Since the Doppler half-width at half-maximum, $\Delta$, exceeds the ground-state hyperfine splittings ($\Delta=2\pi\!\times 390$~MHz for \iso{39}{K} at $300$~K), there are also nearby pairs of \textit{ground-state} crossover resonances for each isotope. This type of crossover resonance leads to \textit{reduced} transmission, as the pump beam transfers greater population to the probed ground state via hyperfine pumping. The significant overlap between various transitions makes it extremely difficult to resolve individual features, and off-resonant excitation means cycling or closed transitions are effectively absent.  In practice, the useful features are limited to the composite feature comprising the \iso{39}{K}, $F=2\rightarrow F^\prime=1,2,3$ transitions and the one due to the two \iso{39}{K} ground-state crossovers. For brevity, we will refer to these features as A and B, respectively (see figure~\ref{fig:schematic}).  These two features will be the focus of this work.

Potassium is widely used in experiments with cold atoms due to the existence of both bosonic and fermionic species, and the availability of inexpensive laser diodes at the cooling wavelengths.  Although we are aware of research groups using heterodyne spectroscopy with potassium for laser locking, to our knowledge there are no published examples showing what the resulting spectra should look like or how they may be optimized.  It is therefore our aim to provide a practical study of these methods for people working in this field. As our interest is in laser stabilization, we attempt to optimize the slope of the discriminant --- \textit{i.e.} the rate of signal variation in V/MHz around the zero-crossing.  This sets the gain for feedback control.  In contrast to electronic gain, there is no inherent compromise between optical gain and total loop bandwidth, making it beneficial to maximize the slope optically.  Despite the overlap of multiple transitions, we will show that narrow ($< 10$ MHz) dispersive features can still be observed.

The rest of the paper is organized as follows.  In section \ref{sec:SatAbs} we study conventional saturated absorption, and examine the feature amplitudes and widths as functions of vapour pressure.  Perhaps surprisingly, the apparent half-width of saturated absorption features can decrease with increasing vapour pressure.  This is shown to be due to the departure from the linear absorption regime.  In section \ref{sec:DM}, we investigate heterodyne detection using direct modulation of the probe.  The signals we obtain are well approximated by derivative signals. In section \ref{sec:MT} we perform modulation transfer spectroscopy by moving the modulation to the pump beam.  This method results in a flatter background and steeper slope for the \iso{39}{K} $F=2\rightarrow F^\prime=3$ feature, at the expense of eliminating the crossover.  For modulation transfer, varying the pump and probe powers reveals information regarding the physical origins of the signal.  Finally, in section \ref{sec:Discussion}, we compare the relative benefits and drawbacks of the two methods, and suggest some ways that they could be improved.

\section{\label{sec:SatAbs}Saturated Absorption}

We begin with conventional saturated absorption spectroscopy \cite{Han71}.  Our experiments use a home-built external-cavity diode laser loosely based on the design presented in \cite{thesisPapp}.  The diode is anti-reflection coated, with a centre wavelength of 770 nm (Eagleyard, EYP-RWE-0790-04000-0750-SOT01-0000\footnote{Product names and part numbers are used for identification purposes only, and do not constitute an endorsement by the authors or their institution.}). Measurements with a home-built fibre ring resonator set an upper limit of $\sim 1$~MHz for the laser linewidth. The laser frequency is scanned by moving a diffraction grating on a kinematic mirror mount with a piezo-electric transducer. The elliptically elongated beam emitted from the laser is made approximately circular with an anamorphic prism pair before passing through a Faraday isolator to prevent feedback to the laser.  Measurements of the power transmitted past a razor on a translation stage give $1/e^2$ intensity radii of $w_x=1.18\pm0.03$~mm and $w_y=1.15\pm0.02$~mm in the horizontal and vertical directions, respectively, at a distance of $0.4$~m from the laser (uncertainties are from fits only).  Similar measurements at 4~m give $w_x=1.59\pm0.03$~mm and $w_y=1.506\pm0.006$~mm. All spectra presented here were obtained at intermediate distances.

\begin{figure}
	\centering
		\includegraphics[scale=0.28,angle=270]{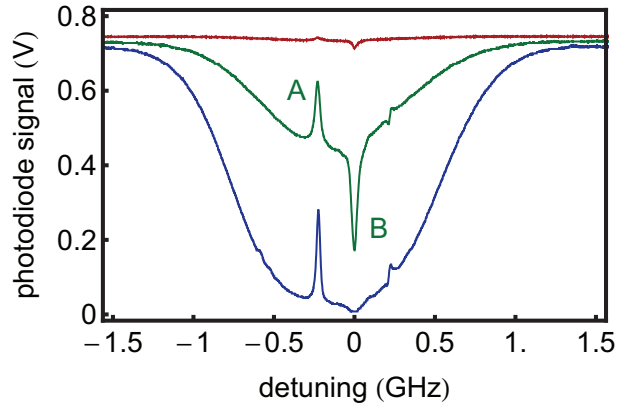}
	\caption{(Colour online.) Saturated absorption at varying cell temperature.  From top to bottom, the curves correspond to temperatures of \degC{21}, \degC{50}, and \degC{67}.}
	\label{fig:satAbsTemp}
\end{figure}

Spectra were taken with a 75~mm Pyrex reference cell (Thorlabs, CP25075-K), containing potassium isotopes in their natural abundance. Because of the low vapour pressure of potassium at room temperature, we began by heating the cell and recording saturated absorption spectra at various temperatures.  The reference cell was wrapped in a flexible resistive heater (Minco HK5464R14.6L12A) and a layer of thermal insulation, and the temperature was monitored with a sensor (Minco S665PDZ40B) pressed between the heater and the cell.  Example spectra are shown in figure~\ref{fig:satAbsTemp}.  Because the isotope shift is smaller than the Doppler shift, and due to the preponderance of \iso{39}{K}, only a single Doppler-broadened background is evident in each scan. This background absorption varies strongly with relatively small changes in temperature.  At \degC{50} the A and B features are both clearly visible.  It is also possible to see the conglomeration of \iso{39}{K}, $F=1\rightarrow F^\prime$ transitions, which nearly overlap with the \iso{41}{K} ground-state crossovers, causing a slightly dispersive shape near 230~MHz.

In order to quantify the temperature dependence, we take alternating scans with and without the pump beam.  By subtracting a purely Doppler-broadened spectrum from one with a saturating beam present, we obtain a single trace with a flat background and narrow peak (dip) for the A (B) feature.  For simplicity we fit the features to Lorentzian functions and extract the amplitudes and widths (HWHM).  The frequency scale throughout this work was calibrated by comparing the difference between the A and B line centres with a Fabry-Perot spectrum analyser with variable mode spacing \cite{Bud00}.  This yielded a frequency difference between A and B features of 224~MHz, with a 4~MHz uncertainty estimated from variations over time and under different conditions (\textit{e.g.}, cavity length, cell temperature, pump and probe intensities).

We model the temperature dependence as follows.  For a weak (\textit{i.e.}, non-saturating) probe, the transmitted signal, normalized to the incident intensity, is given by $\exp(-\alpha)$ where $\alpha$ is the absorption coefficient.  For a two-level system, with $\Delta\gg\gamma$, the saturating beam leads to a Lamb dip in the probe absorption coefficient \cite{Pap80,Ber11}
\begin{eqnarray}
\label{eq:alphaS}
\alpha_S &=& \alpha_D\,(1-\mathcal{L})
\end{eqnarray}
where $\alpha_D$ is the Doppler-broadened absorption coefficient, and $\mathcal{L}=a_L/(1+\delta^2/\Gamma^2)$ is a Lorentzian function of detuning $\delta$ from the centre of the Doppler-free feature. The amplitude and width of $\mathcal{L}$ are functions of the pump beam intensity.  We allow different centre frequencies for the Doppler background and $\mathcal{L}$ (recall figure~\ref{fig:satAbsTemp}), and take $\alpha_D$ as the value of the Doppler absorption coefficient at the resonance.  Since $\alpha_D$ is proportional to the vapour pressure $\Pvap$, we explicitly write $\alpha_D=a_G\Pvap$.  The difference between transmission with and without the saturating beam is then
\begin{eqnarray}
\label{eq:DeltaS}
\Delta S = e^{-a_G\Pvap}\left[e^{a_La_G\Pvap/(1+\delta^2/\Gamma^2)}-1\right]
\end{eqnarray}
The amplitude and width are now easily obtained, using the dependence of $\Pvap$ on temperature described in \cite{Alc84}.  For the B feature, we simply take $a_L<0$ to obtain a negative feature.

\begin{figure*}~\\
	\centering
		\parbox{0.5\textwidth}{\includegraphics[angle=270,width=0.9\linewidth]{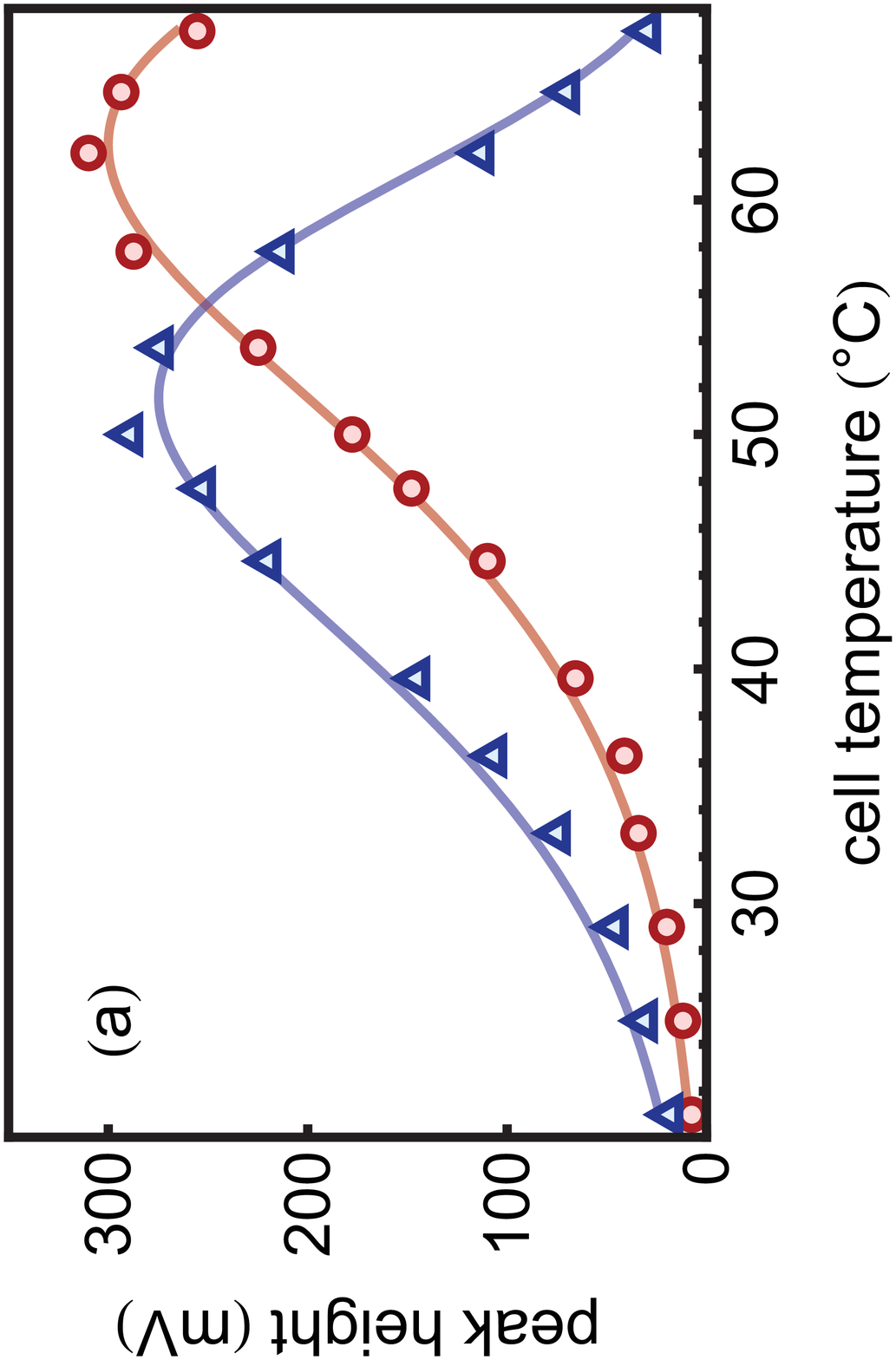}}\hfill%
		\parbox{0.5\textwidth}{\includegraphics[angle=270,width=0.9\linewidth]{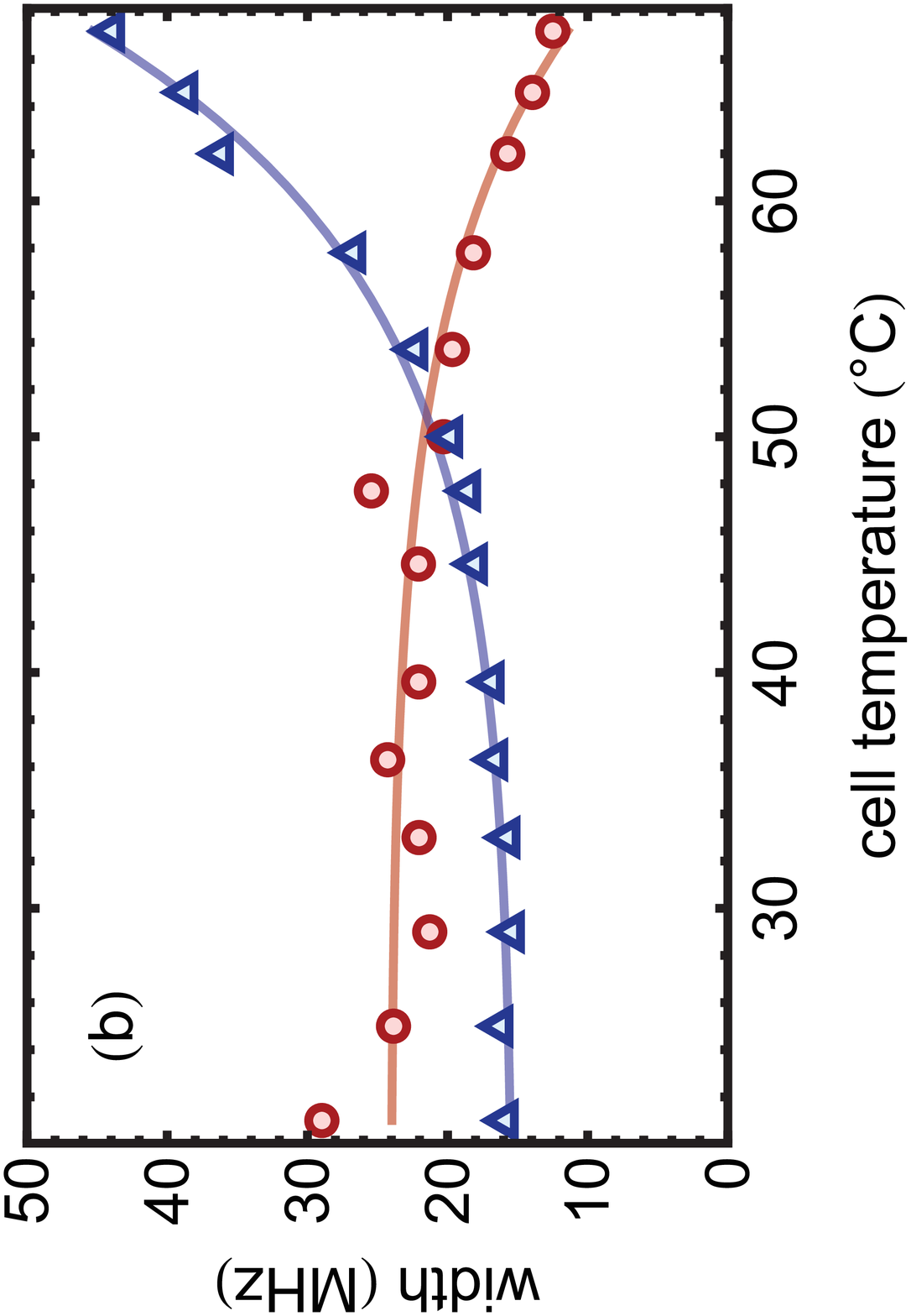}}
	\caption{(Colour online.) Saturated absorption amplitudes and widths with varying cell temperature.  The pump power was $1.19$~mW and the probe was $0.23$~mW (both measured immediately before the cell), corresponding to $20\pm 4$~mW/cm$^2$ and $3.9\pm 0.7$~mW/cm$^2$, averaged over the Gaussian beam profile. (a) Amplitude. Red circles are for the A feature (due to \iso{39}{K}, $F=2\rightarrow F^\prime$ transitions), and blue triangles the B feature (due to the \iso{39}{K} ground-state crossovers).  Solid curves are fits derived from (\ref{eq:DeltaS}). (b) Half-width at half-maximum.}
	\label{fig:satAbsHWTemp}
\end{figure*}

The observed dependence of the feature amplitudes and widths on cell temperature are shown in figure~\ref{fig:satAbsHWTemp}, together with the predictions of (\ref{eq:DeltaS}). The agreement between experiment and theory is excellent.  The crossover (B) amplitude peaks first, around \degC{50}, but is then compressed by the deep Doppler absorption; feature A is largest at \degC{62}.  Note that these temperatures depend on the length of the cell.  Both features have similar maximum amplitudes for these data, but this may depend in general on the intensities of the pump and probe beams.  The observed linewidths are larger than the natural width $\sim 3$~MHz of a single transition.  However, due to the composite nature of each feature, this is not too surprising. The A feature comprises three transitions distributed over 30~MHz, and the B feature is due to a pair of transitions $9.4$~MHz apart.  Including power broadening, this is in reasonable agreement with our observations. Since our model neglects temperature-dependent broadening mechanisms such as collisions and finite transit times, the observed variations in width can be attributed solely to the exponential dependence of transmitted signal on the absorption coefficients.

Based on the results of figure~\ref{fig:satAbsHWTemp}, we typically keep our cell around 46--\degC{48}, corresponding to a maximum Doppler-broadened absorption of 35--40\% for a weak probe.  This is sufficient to produce a strong signal at each feature, with minimal experimental complexity.

\section{\label{sec:DM}Direct Probe Modulation}

\begin{figure}~\\
	\centering
		\parbox{0.5\textwidth}{\includegraphics[width=0.9\linewidth]{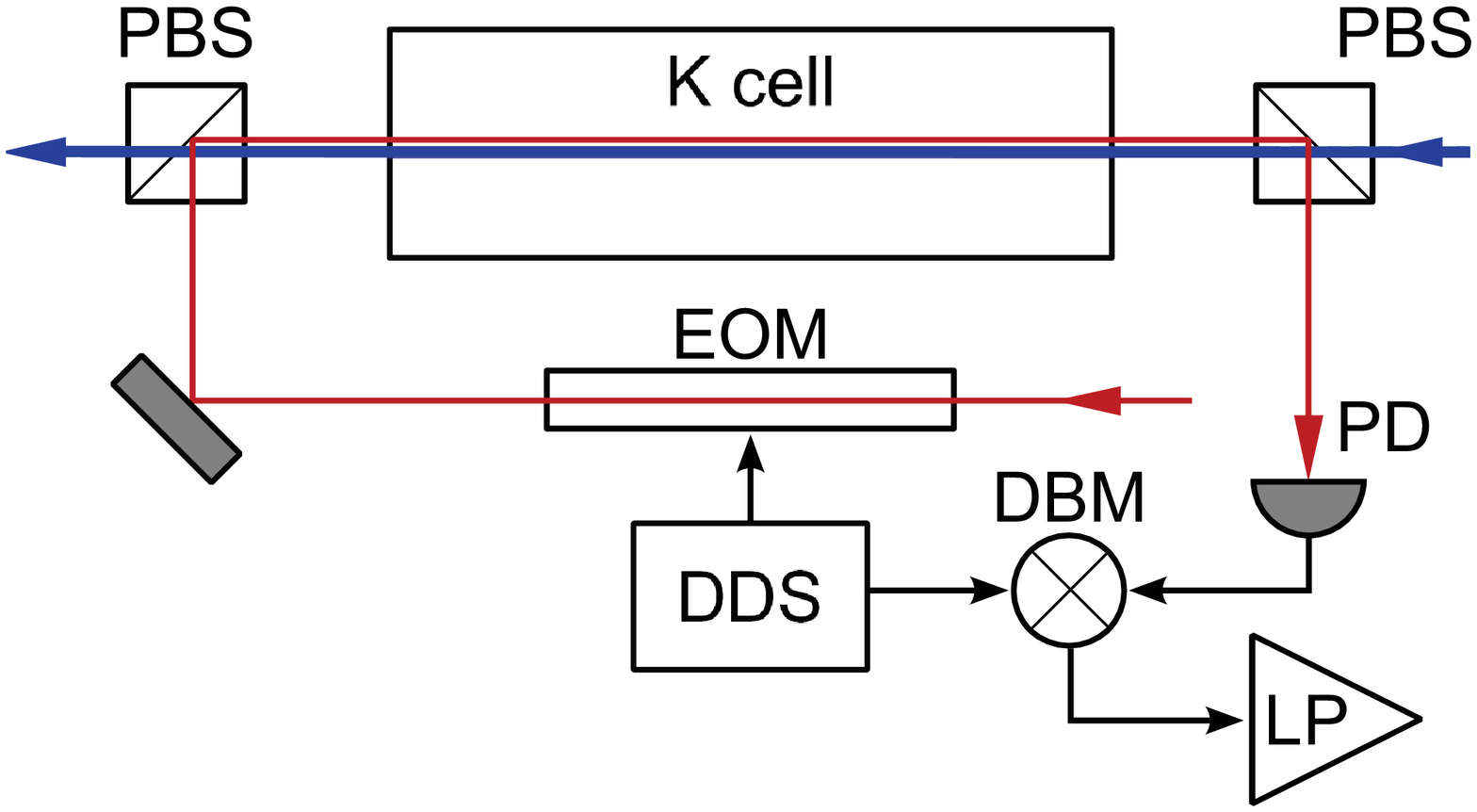}}\hfill%
		\parbox{0.5\textwidth}{\includegraphics[angle=270,origin=B,width=0.9\linewidth]{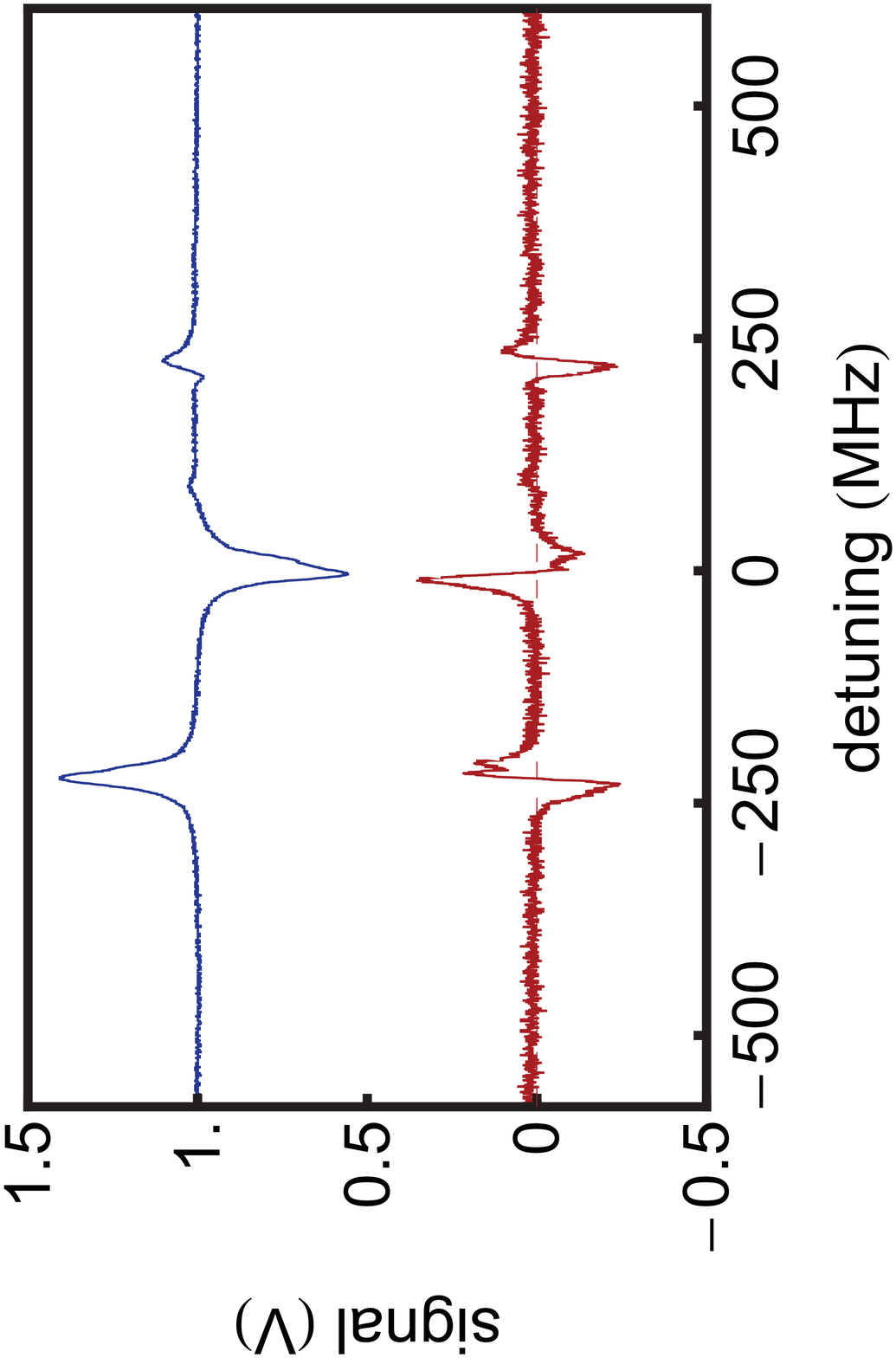}}
	\caption{(Colour online.) Direct modulation spectroscopy.  Left: simplified schematic.  The heavy blue line shows the pump, and the thin red line the probe. The beams are shown separated for clarity, but overlap in the experiment. PBS: polarizing beam splitter; EOM: electro-optic modulator; DDS: direct digital synthesis; PD: photodiode; DBM: double-balanced mixer; LP: low-pass filter.  Right: comparison of direct modulation and saturated absorption. The lower trace is the mixed-down signal (\textit{i.e.}, the output of the low-pass filter, LP), and the upper one the subtracted saturated absorption signal.  (The upper curve has been shifted vertically for clarity.)  The powers were $50\,\mu$W for the probe and $250\,\mu$W for the pump, corresponding to $0.85\pm 0.16$~mW/cm$^2$ and $4.2\pm 0.8$~mW/cm$^2$.}
	\label{fig:FMvSatAbs}
\end{figure}

We now turn our attention to heterodyne detection, beginning with direct phase modulation of the probe beam. Our optical setup was similar to the one presented in \cite{McC08}; a simplified schematic is shown in figure~\ref{fig:FMvSatAbs}. The counter-propagating pump and probe beams had orthogonal linear polarizations. To obtain the phase modulation, we used a home-built electro-optic modulator (EOM).  The LiNbO$_3$ crystal (Casix Optics) was Y-cut, measuring $2\times2\times25$~mm, with anti-reflection coating on the input and output faces, and gold coating on the sides (Z-faces) to form electrodes. An operational amplifier and resonant resistor-inductor-capacitor (RLC) circuit were used to drive the crystal at $9.62$~MHz. This frequency was chosen to be about half the linewidths observed in figure~\ref{fig:satAbsHWTemp}(b), as a compromise between obtaining a large slope and maintaining a derivative lineshape \cite{Bjo83,Bla01}. The modulation was generated by a four-channel direct digital synthesis evaluation board (Analog Devices, AD9959/PCBZ-ND) which allowed adjustment of the relative phase between the local oscillator (LO) and the radio-frequency (RF) signal from the photodiode during demodulation. The transmitted probe beam was detected with an amplified silicon photodiode having $100\,\mathrm{k}\Omega$ transimpedance gain and 50~MHz bandwidth (Thorlabs PDA8A/M). A bias-T (Mini-Circiuts ZFBT-4R2GW) was used to separate the low- and high-frequency components of the signal.  The low-frequency port was fed directly to an oscilloscope, allowing us to observe saturated absorption spectra.  The high-frequency component was amplified (Mini-Circuits ZFL-500LN), mixed down with a double-balanced mixer (Mini-Circuits ZRPD-1), and filtered with a home-built active low-pass filter ($4^{\rm th}$-order Butterworth) with cutoff frequency $200$~kHz and a gain of 2.  The LO power was $+7$~dBm.  We found that higher LO powers increased the noise without significantly improving the signal, while lower powers resulted in nonlinear behaviour of the phase detector at low RF power.

An example of a typical signal obtained \textit{via} direct probe modulation is shown in figure~\ref{fig:FMvSatAbs}.  The signal is compared to a background-subtracted saturated absorption spectrum for reference.  As expected, the modulation signal looks like the derivative of the absorption.  At these lower pump and probe powers, a number of features which were obscured in figure~\ref{fig:satAbsTemp} can now be seen.  For example, the internal structure of the A and B transitions leads to kinks in the modulation signal.  This is especially problematic for the crossover, as the asymmetry pushes the kink towards the zero-crossing lock point.  We will return to this point below. As mentioned previously, the feature to the blue of the B feature, is due predominantly to the \iso{39}{K}, $F=1\rightarrow F^\prime$ transitions (there is a small contribution of opposite sign due to the \iso{41}{K} ground-state crossovers).  This feature leads to a distorted zero crossing and saturates at much lower powers than the A and B features.  We therefore neglect it in the rest of our analysis.

\begin{figure*}~\\
	\centering
		\parbox{0.5\textwidth}{\includegraphics[angle=270,width=0.9\linewidth]{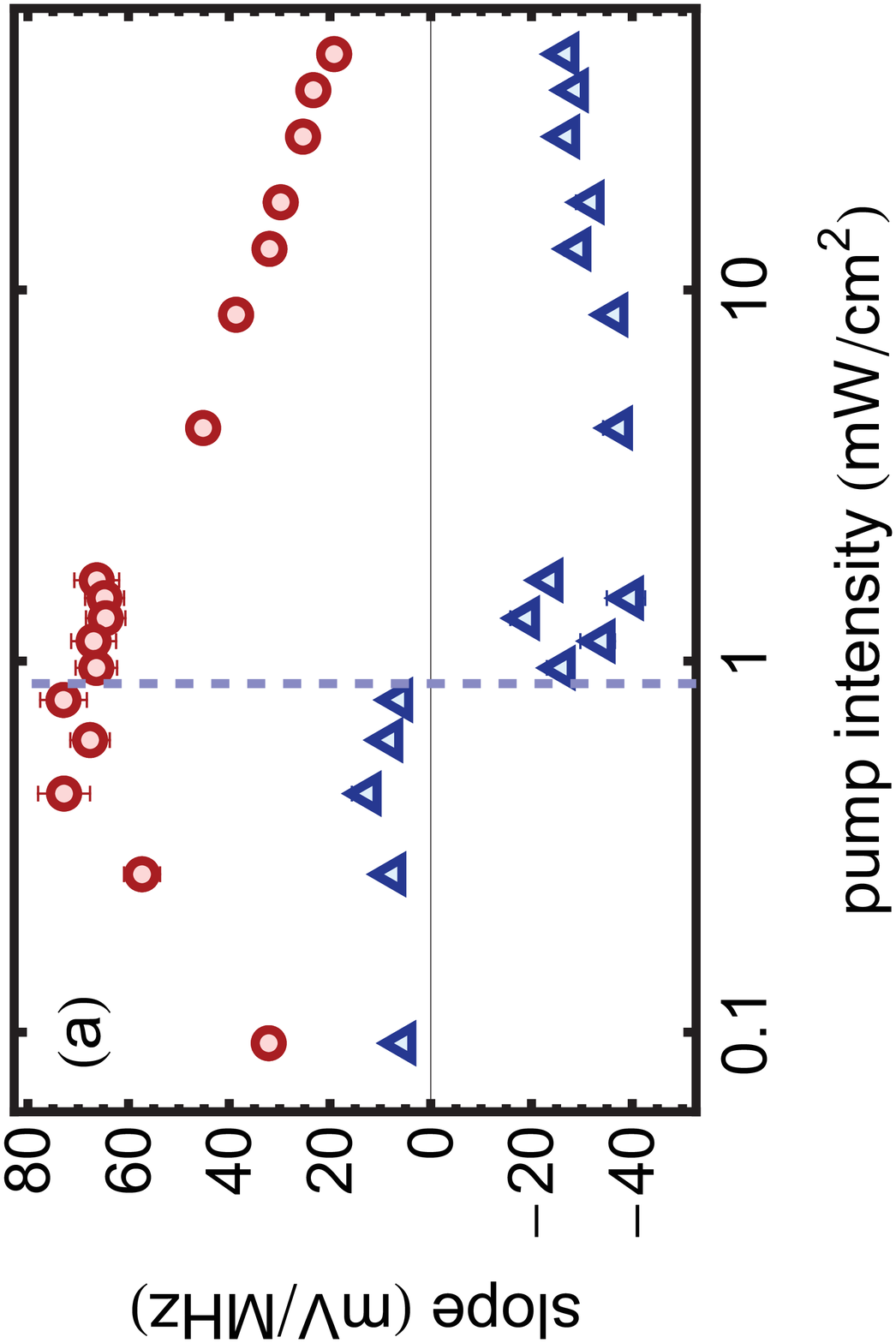}}\hfill%
		\parbox{0.5\textwidth}{\includegraphics[angle=270,width=0.9\linewidth]{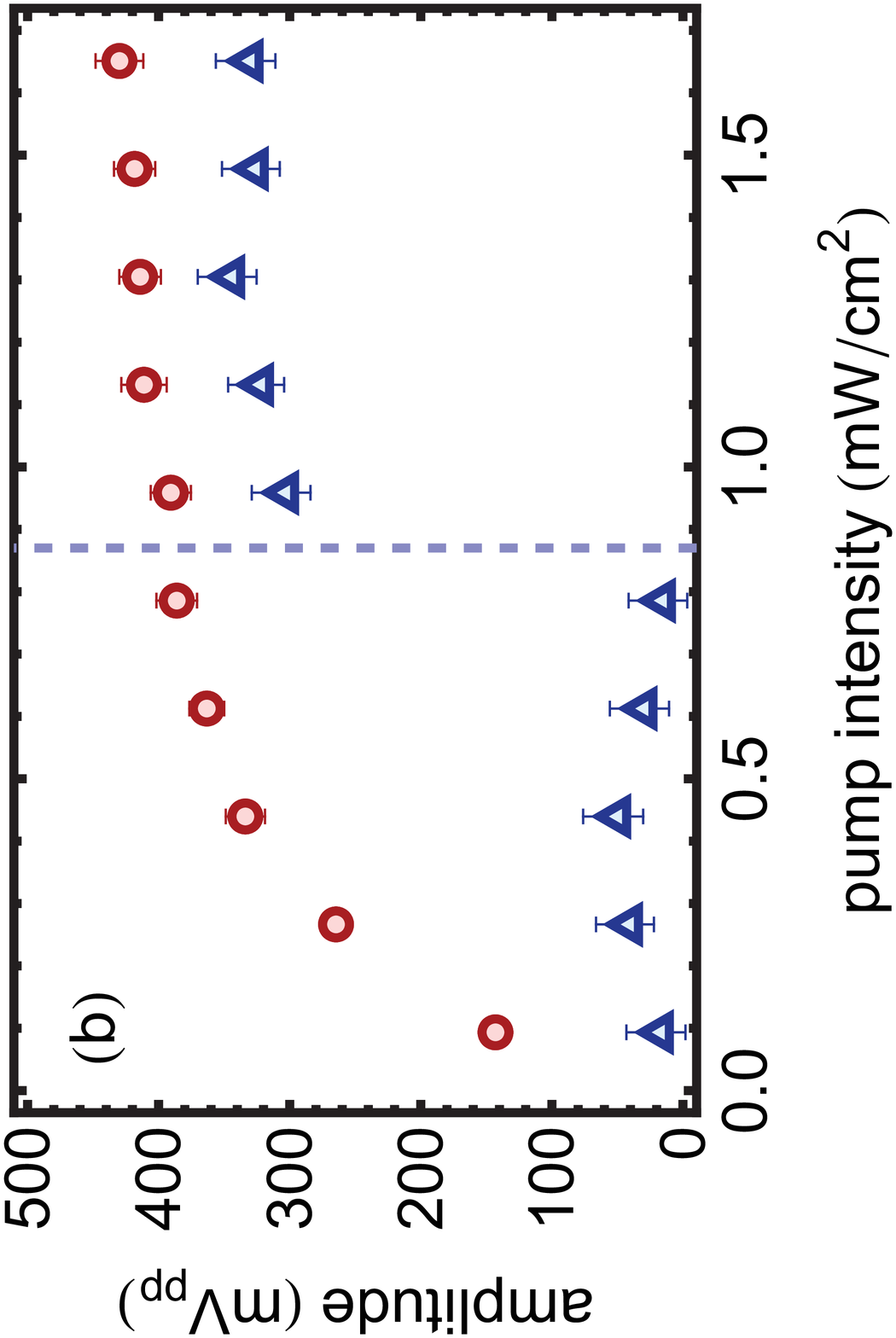}}
	\caption{(Colour online.) Dependence of direct modulation spectra on pump intensity, with fixed $0.85\pm 0.16$~mW/cm$^2$ probe intensity. (a) Slope at the zero-crossing.  The red circles (blue triangles) are data for the A (B) feature. The vertical dashed line shows when the kink in the B feature no longer hits the zero crossing.  (b) Peak-to-peak amplitude, between turning points.  Here and throughout this work, error bars reflect fit uncertainties only; when not visible, they are smaller than the plot markers. Not shown is an approximately $20\%$ systematic uncertainty in intensity, due to variation in the Gaussian spot size away from the laser.}
	\label{fig:FMPump}
\end{figure*}

The behaviour of the direct modulation signal as a function of pump intensity is shown in figure~\ref{fig:FMPump}. The slope at the zero crossing of the A feature increases until the saturated absorption amplitude begins to saturate and the width broadens, causing the derivative signal to roll over.  The result is an optimum slope for the A feature of $\sim 70$~mV/MHz just below 1~mW/cm$^2$.  The B feature slope is complicated by the kink which occurs at low intensities, when the two crossover transitions (separated by $9.4$~MHz) are partially resolved.  At $50\,\mu$W ($0.85\pm 0.16$~mW/cm$^2$) the kink moves below zero, and then is washed out completely as the pump power increases (dashed blue line).  The amplitude shows this transition more dramatically. The A feature amplitude saturates at much lower intensity than we expected.  The saturation intensity for the cycling transition is $1.75$~mW/cm$^2$ \cite{Wan97,Fal06}.  Assuming isotropic populations of the atomic Zeeman substates, this increases to $3.75$~mW/cm$^2$.  Additional relaxation mechanisms (\textit{e.g.}, de-phasing or transit time) tend to further increase the saturation intensity \cite{Pap80}.  We speculate that optical pumping may play a role in our observations.  The importance of hyperfine pumping has been previously demonstrated for rubidium \cite{Smi04}, and the reduced level splittings in potassium could exacerbate this effect by increasing the of-resonant excitation to the non-cycling transitions. As our aim here is simply to optimize the locking signal, we leave the origins of this effect for future study.

\begin{figure*}~\\
	\centering
		\parbox{0.5\textwidth}{\includegraphics[angle=270,width=0.9\linewidth]{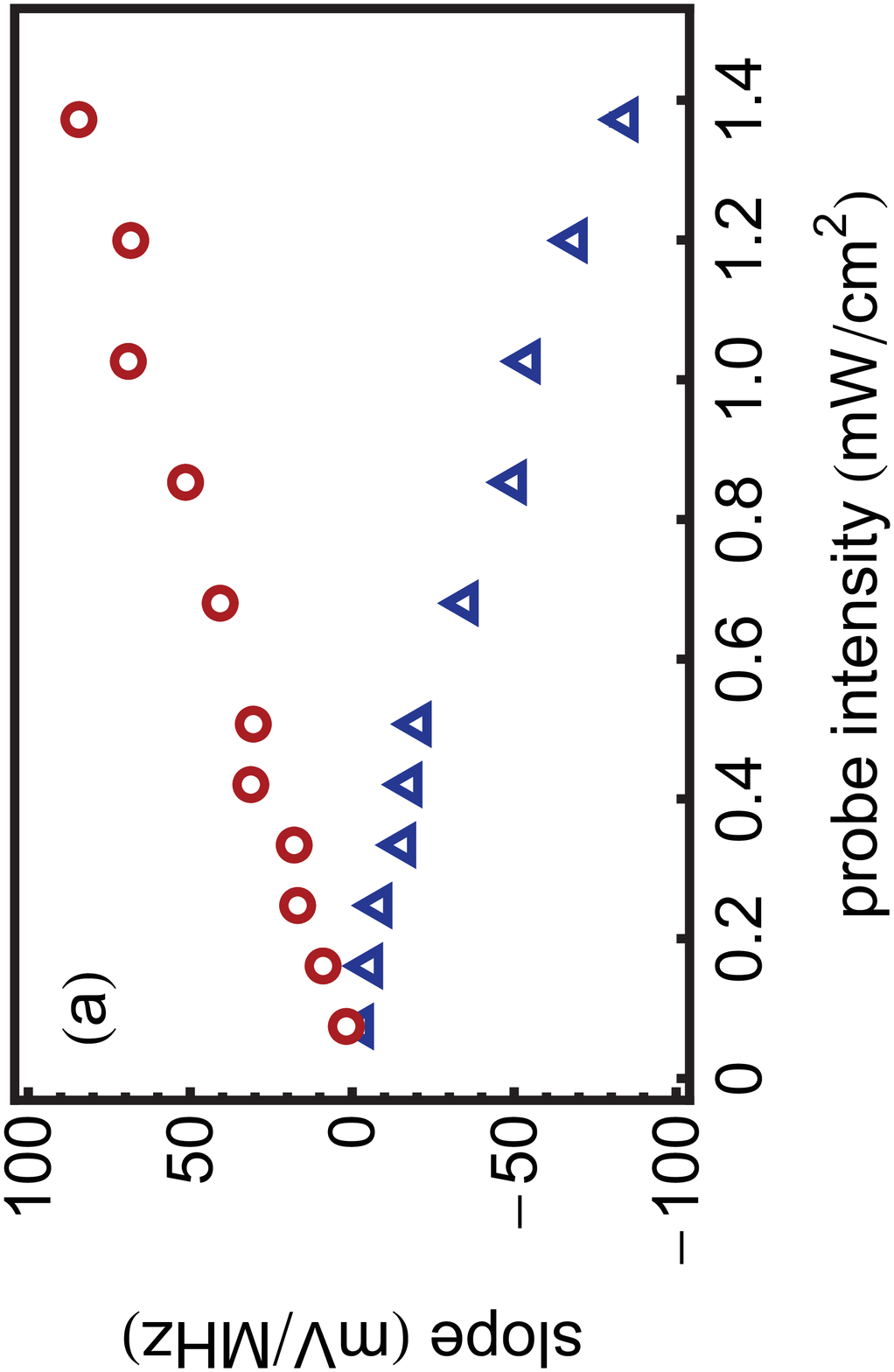}}\hfill%
		\parbox{0.5\textwidth}{\includegraphics[angle=270,width=0.9\linewidth]{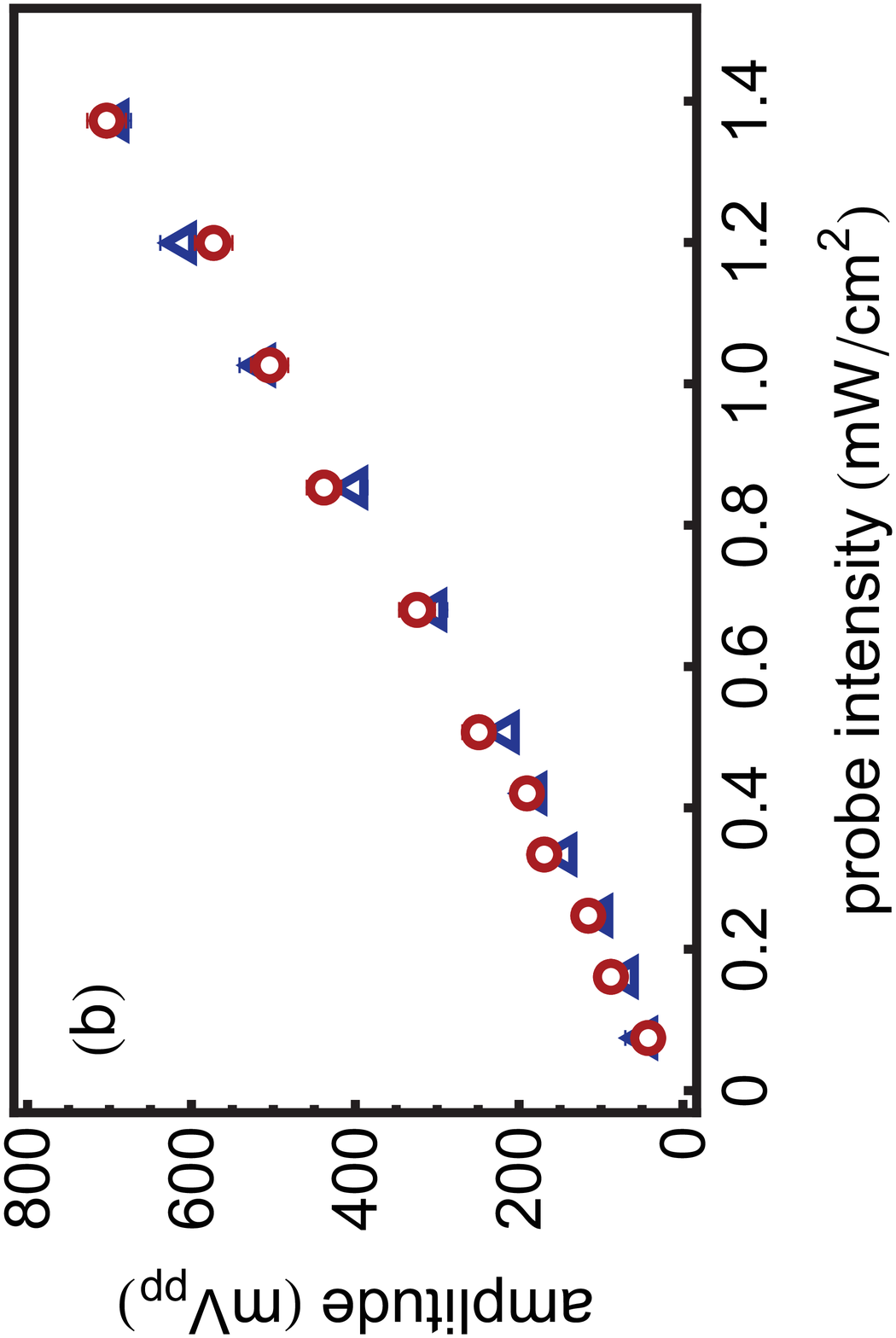}}
	\caption{(Colour online.) Dependence of direct modulation spectra on probe intensity, with fixed $4.2\pm 0.8$~mW/cm$^2$ pump intensity. (a) Slope at the zero-crossing.  Points are as in figure~\ref{fig:FMPump}. (b) Peak-to-peak amplitude.}
	\label{fig:FMProbe}
\end{figure*}

Figure \ref{fig:FMProbe} shows the effect of varying the probe intensity. For this choice of pump intensity ($4.2\pm 0.8$~mW/cm$^2$), the two features show similar slopes and amplitudes.  For the highest probe intensities we have tested (limited by beam size and saturation of the photodiode), we do not see any clear saturation.

\section{\label{sec:MT}Modulation Transfer}

\begin{figure*}~\\
	\centering
		\parbox{0.5\textwidth}{\includegraphics[width=0.9\linewidth]{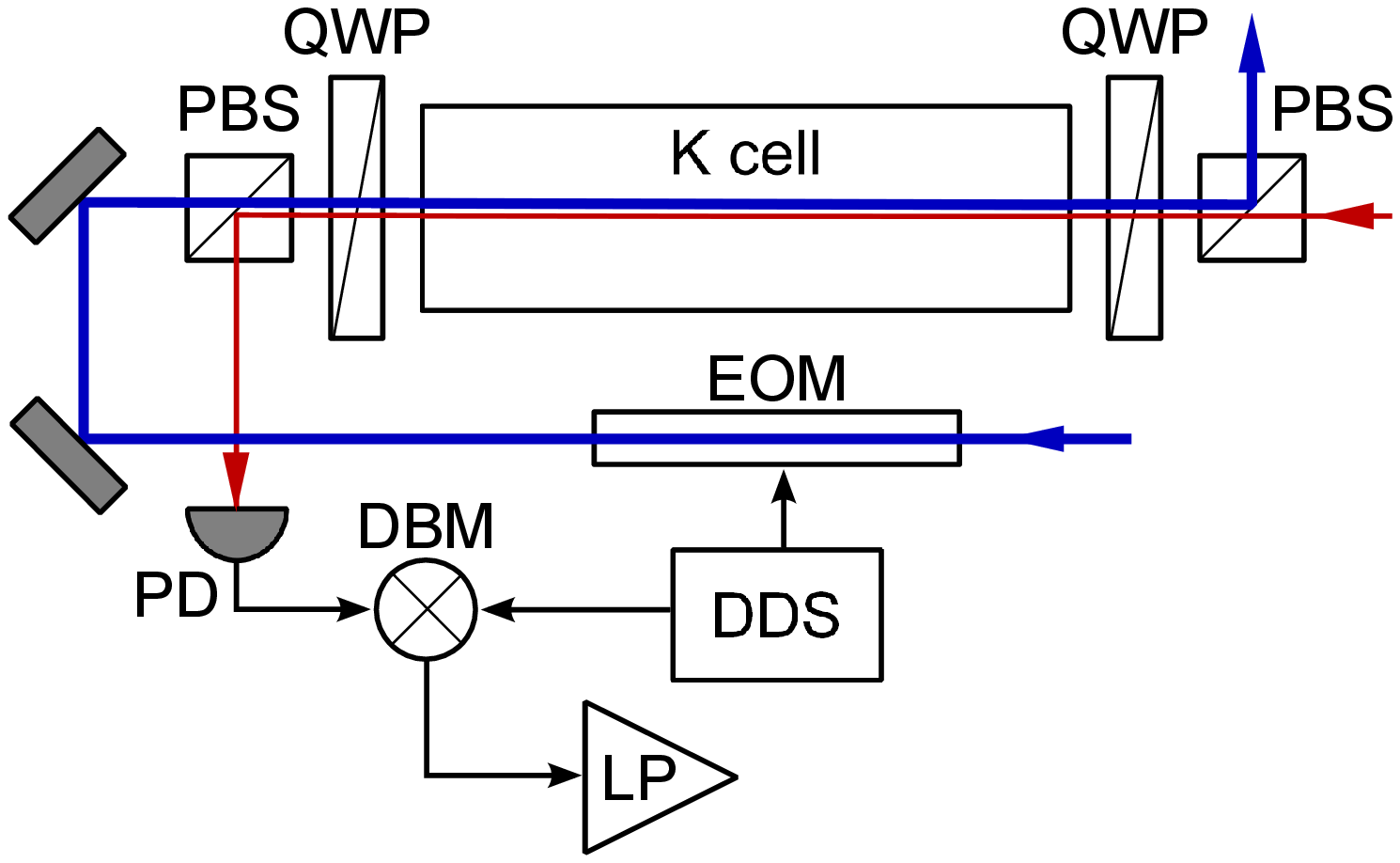}}\hfill%
		\parbox{0.5\textwidth}{\includegraphics[angle=270,origin=B,width=0.9\linewidth]{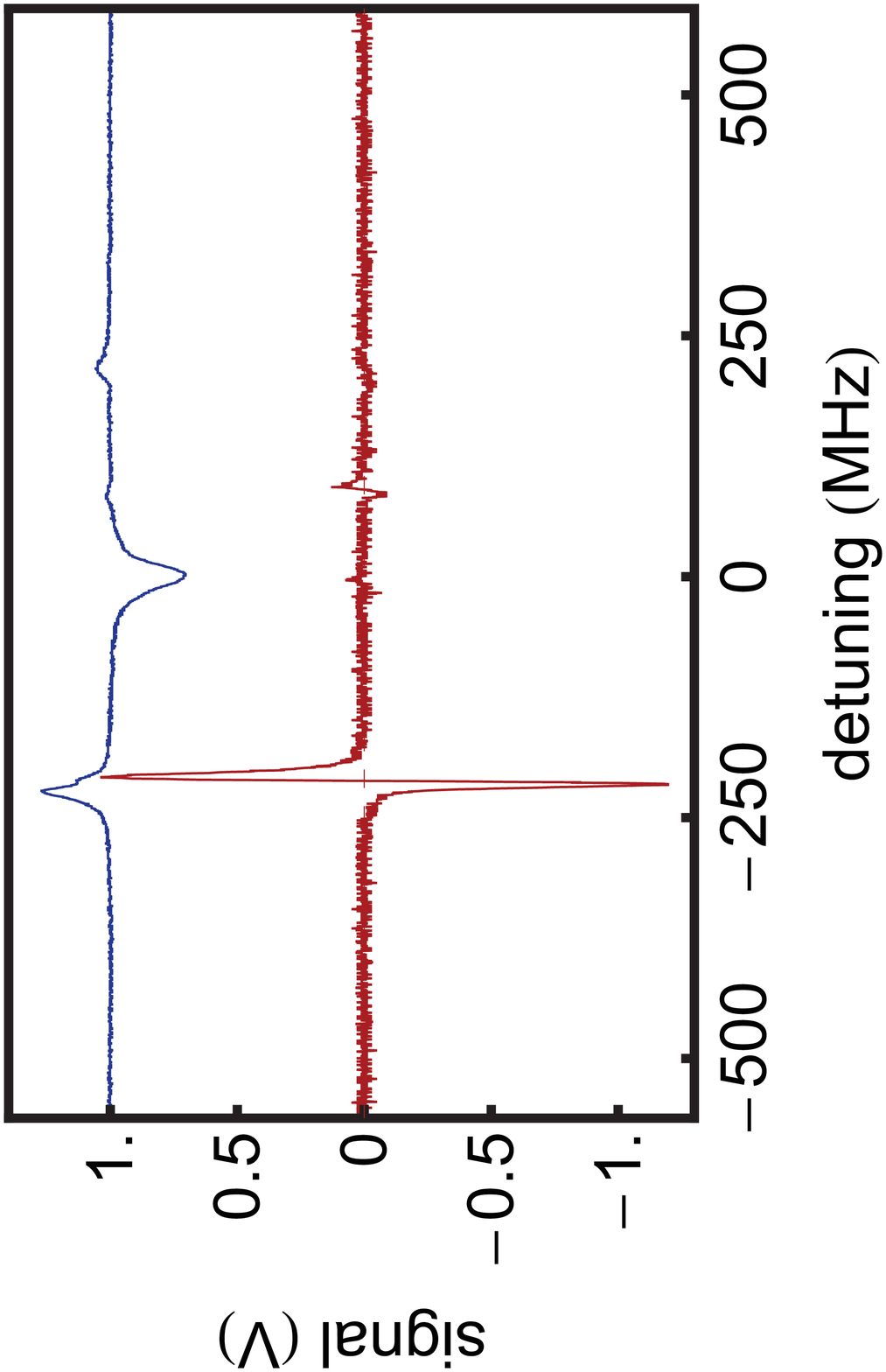}}
	\caption{(Colour online.) Modulation transfer spectroscopy.  Left: simplified schematic.  QWP: quarter-wave plate; all other notations are as in figure~\ref{fig:FMvSatAbs}.  Right: comparison with saturated absorption.  Below is the heterodyne signal, and above is the subtracted saturated absorption signal (shifted vertically for clarity.) The pump and probe powers are the same as in figure~\ref{fig:FMvSatAbs}.}
	\label{fig:MTvSatAbs}
\end{figure*}

If the phase modulation is moved from the probe to the saturating beam, we obtain modulation transfer spectra, shown schematically in figure~\ref{fig:MTvSatAbs}.  We found that modulation transfer worked better at lower frequency.  An example is shown in figure~\ref{fig:MTvSatAbs}, with the modulation at $1.82$~MHz.  This frequency was chosen based on results in other labs with rubidium \cite{McC08,Neg09}.  We could not achieve a high-$Q$ RLC circuit at this frequency, apparently due to the parasitic properties of available inductors.  We therefore replaced the op-amp driving the EOM with a higher power amplifier (Amplifier Solutions Corporation ASC2832).  We also found that the signal amplitude more than doubled if we used $\sigma^+$-$\sigma^+$ polarizations (\textit{i.e.}, right-hand circular probe and left-hand circular pump), at the expense of nearly complete suppression of the already small B feature.  Polarization-dependent effects have been investigated for linear pump and probe polarizations in $^{87}$Rb \cite{Zha03,Noh11}, but to our knowledge the $\sigma^+$-$\sigma^+$ configuration has not been studied.  No enhancement was observed in our experiments with direct modulation.

The modulation transfer spectrum in figure~\ref{fig:MTvSatAbs} differs from the direct modulation spectrum (figure~\ref{fig:FMvSatAbs}) in a few important ways.  As mentioned above, there is virtually no B feature, and the A feature is noticeably steeper, and with flatter background.  The locking signal is well described by the derivative of a single Lorentzian, with observed widths (HWHM) as small as $4.51\pm0.11$~MHz. These charactersitics are all attributed to the nonlinear nature of the processes generating the probe sidebands.  Modulation transfer spectroscopy is typically strongest for cycling transitions \cite{McC08,Zha03,Noh11}, which allow the inherently weak four-wave mixing of pump and probe fields to occur many times. This also strongly suppresses the Doppler background.

\begin{figure*}~\\
	\centering
		\parbox{0.5\textwidth}{\includegraphics[angle=270,width=0.9\linewidth]{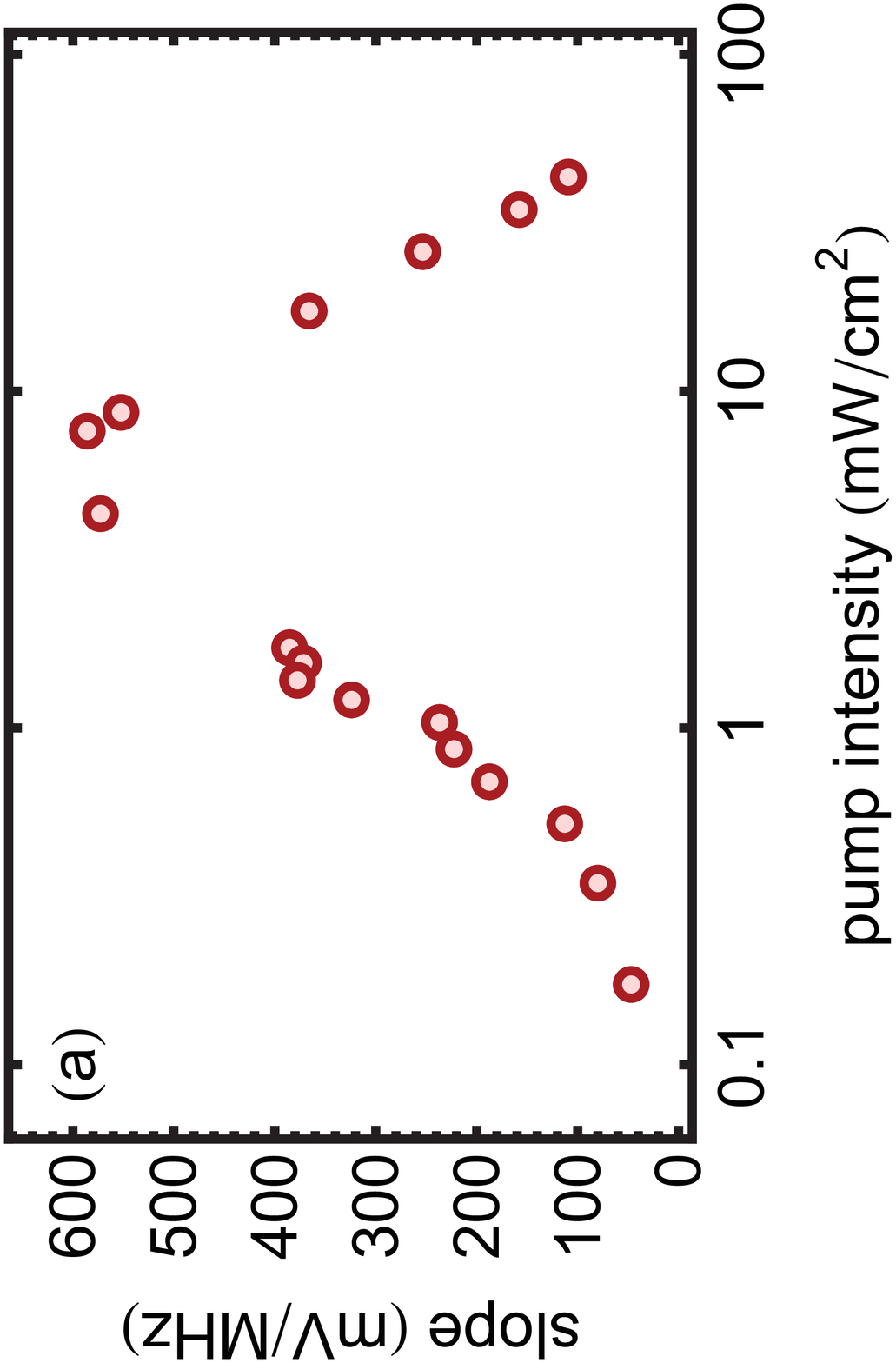}}\hfill%
		\parbox{0.5\textwidth}{\includegraphics[angle=270,width=0.9\linewidth]{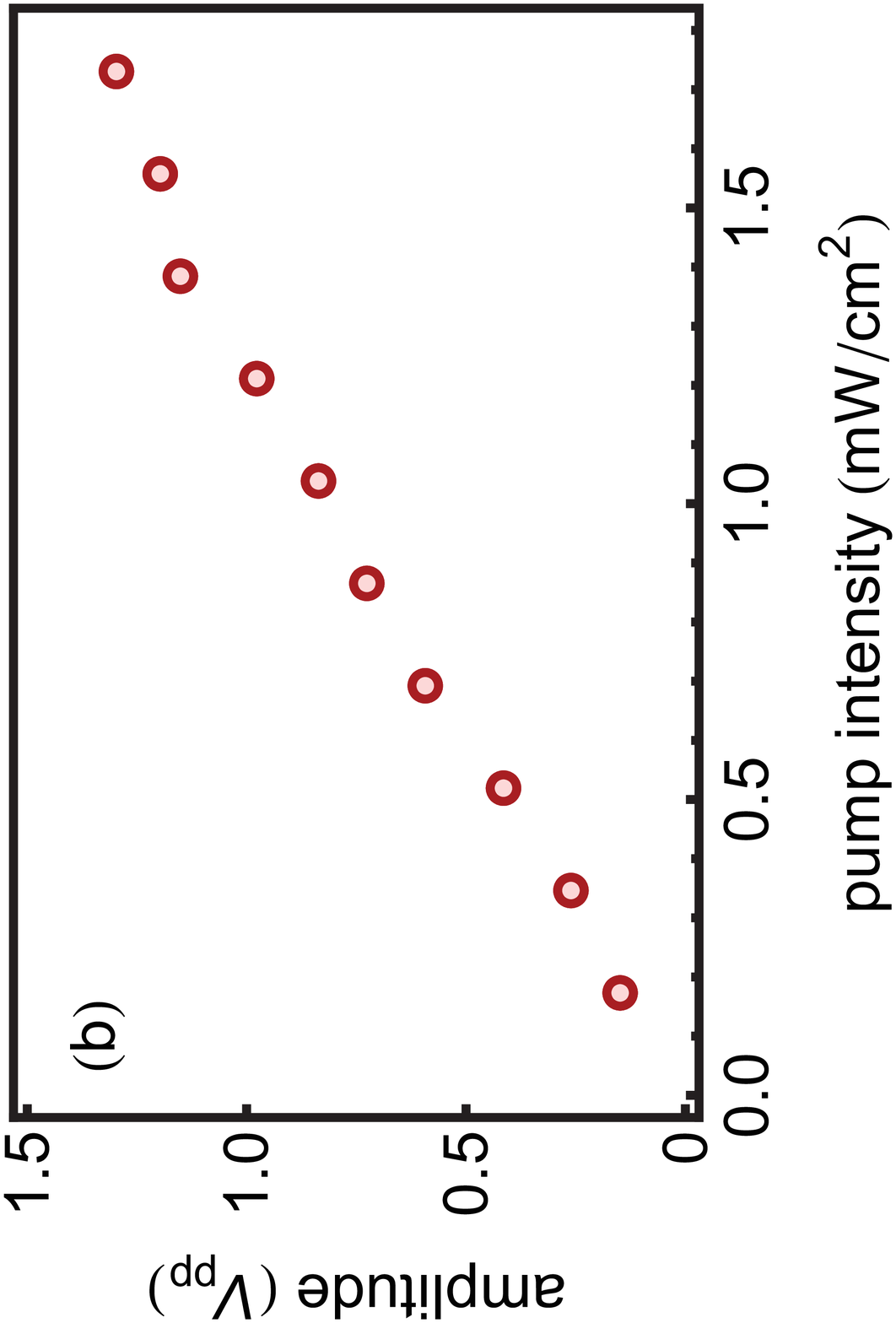}}
	\caption{(Colour online.) Dependence of modulation transfer spectroscopy on pump intensity. The probe intensity was fixed at $0.85\pm 0.16$~mW/cm$^2$.  (a) Slope at the zero crossing. (b) Peak-to-peak amplitude, for small pump intensities.}
	\label{fig:MTPump}
\end{figure*}

Figure \ref{fig:MTPump} shows the dependence of the modulation transfer signal on the pump beam intensity.  The amplitude and slope grow linearly at first, eventually giving way to broadening as in the case of direct modulation.  However, for modulation transfer the optimum pump intensity is much higher and the slope is an order of magnitude greater.

\begin{figure*}~\\
	\centering
		\parbox{0.5\textwidth}{\includegraphics[angle=270,width=0.9\linewidth]{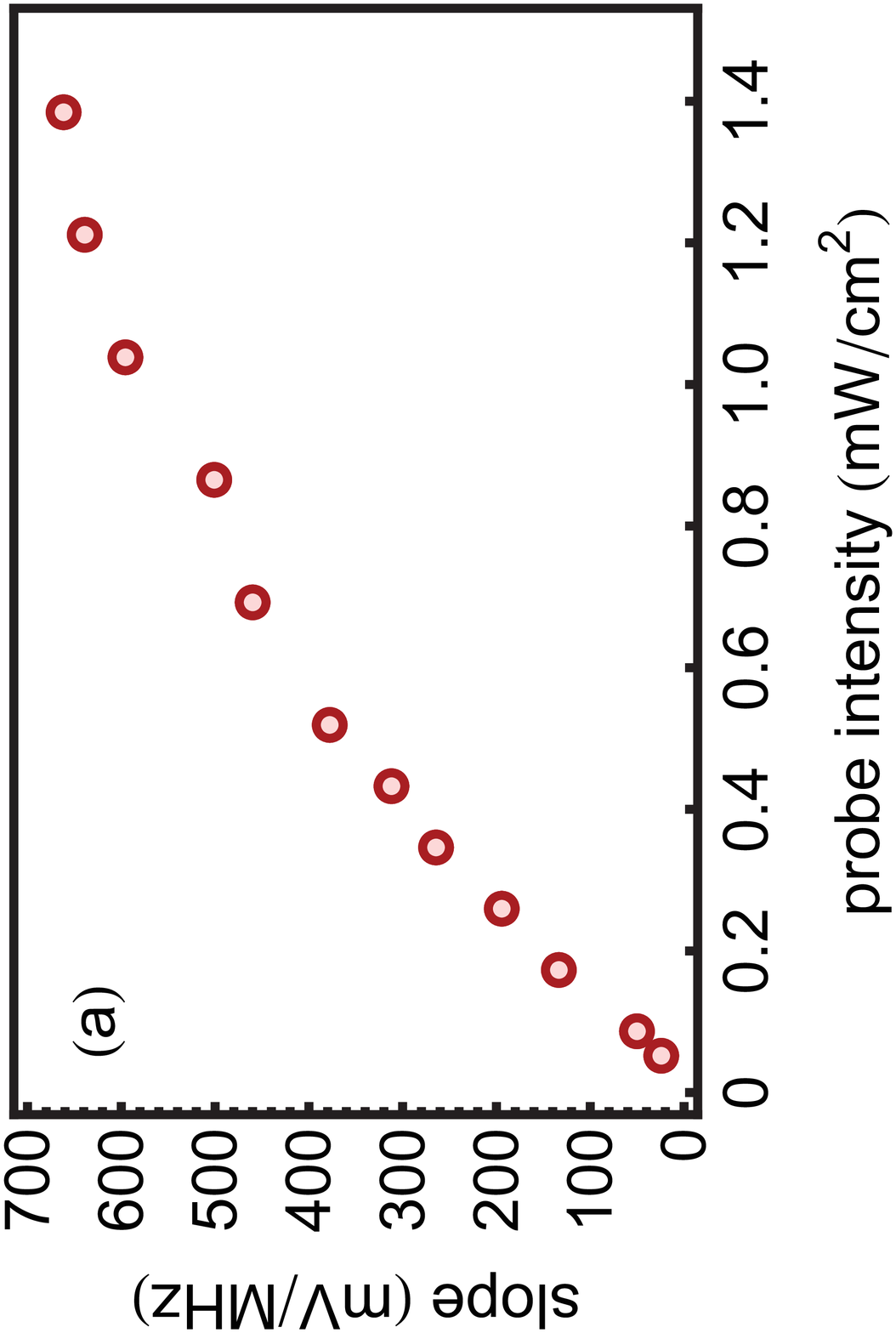}}\hfill%
		\parbox{0.5\textwidth}{\includegraphics[angle=270,width=0.9\linewidth]{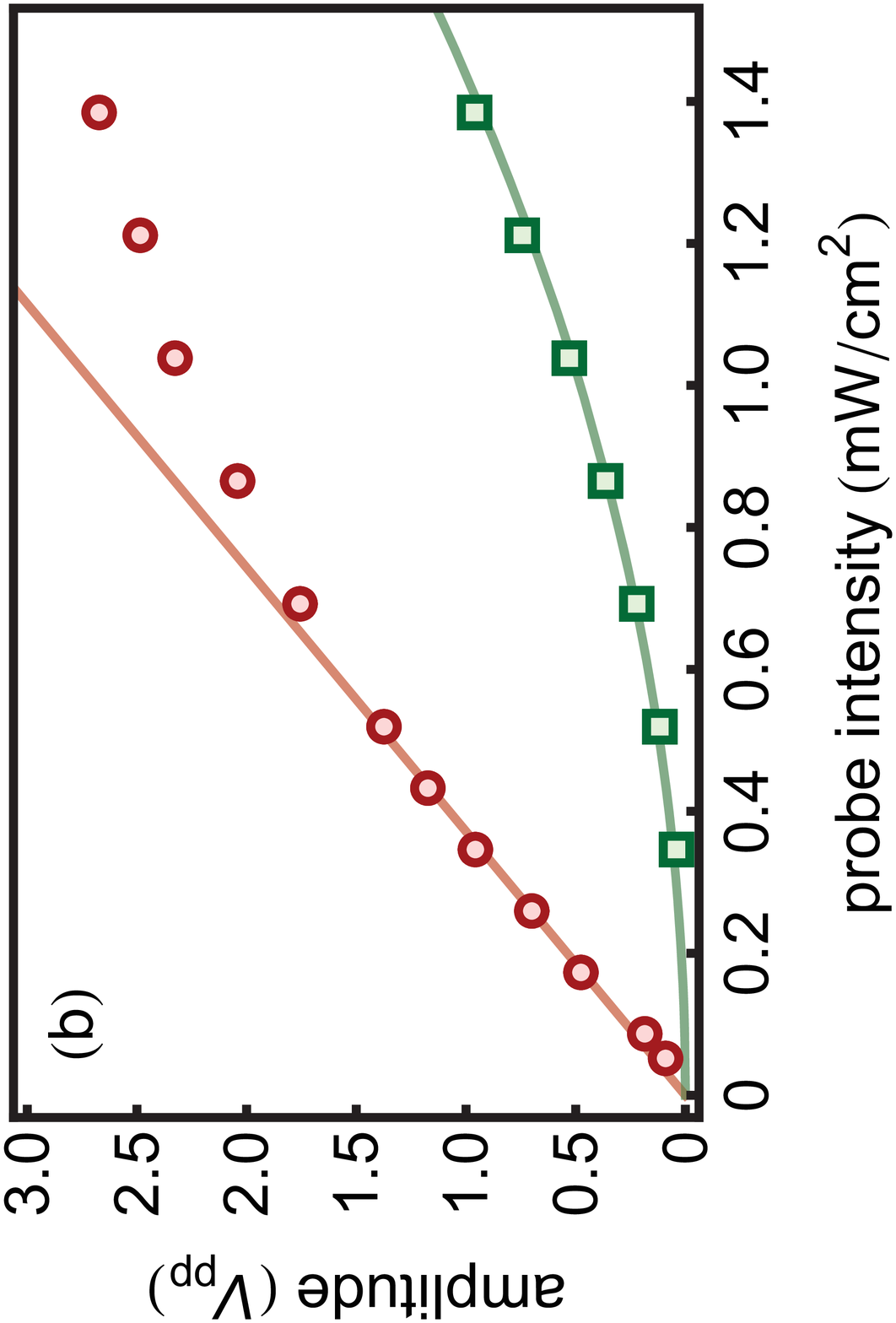}}
	\caption{(Colour online.) Dependence of modulation transfer spectroscopy on probe intensity. The pump intensity was fixed at $4.2\pm 0.8$~mW/cm$^2$.  (a) Slope at the zero crossing, for $1.82$~MHz modulation.  (b) Peak-to-peak amplitude.  Red circles are data with $1.82$~MHz modulation, and green squares with $9.62$~MHz.  The solid red line is a linear fit for low intensities, and the green curve is a power-law fit consistent with quadratic growth.}
	\label{fig:MTProbe}
\end{figure*}

The dependence of the modulation transfer spectrum on the probe intensity is shown in figure~\ref{fig:MTProbe}.  The slope grows linearly and then begins to roll over. The amplitude is also linear for small probe intensity.  Together with the linear dependence on pump intensity in figure~\ref{fig:MTPump}(a), this suggests the physical mechanism of \textit{modulated hole burning} \cite{Shi82}.  The pump carrier and either sideband combine to generate a Lamb dip of oscillating depth.  This generates a probe sideband which beats with the carrier at the detector.  For comparison, we show the data obtained with $9.62$~MHz modulation, together with a fit to a power law giving an exponent $2.01\pm0.05$. This behaviour is characteristic of \textit{reflection}, which can arise in two ways. First, the pump and probe beams burn a standing wave in the atomic ground-state population.  A pump sideband then Bragg-reflects in the backwards direction, creating a sideband on the probe.  Alternatively, a pump sideband can interfere with the probe to form a moving population grating.  In this case the pump carrier is back-reflected and Doppler-shifted, again resulting in a probe sideband. Third-order perturbation theory predicts that these features vanish under thermal averaging when $\Delta\gg\gamma$, but higher-order interactions were shown to result in a signal which is proportional to pump intensity (not shown) and quadratic in probe intensity \cite{Shi82}, consistent with our observations.

\section{\label{sec:Discussion}Discussion}

We have investigated saturated absorption spectroscopy of the D$_2$ lines of potassium, and heterodyne spectroscopy using direct phase modulation of the probe and modulation transfer from the pump.  It was shown that the saturated absorption signal strength can be increased by an order of magnitude or more with moderate heating of the vapour, and the exponential absorption of the probe beam can lead to either an increase or decrease in the apparent half-width of the features.  The comparison between direct modulation and modulation transfer showed that the two methods differ significantly despite the similar experimental arrangements.  The direct modulation signal is roughly proportional to the derivative of the saturated absorption, yielding frequency discriminants for both dominant spectroscopic features.  In contrast, the modulation transfer spectrum showed narrower features with essentially no Doppler background.  Although there was no crossover feature, the maximum slope near the cycling transition was significantly higher than could be obtained by direct modulation.  Investigation of the dependence of pump and probe intensities suggested this feature was dominated by modulated hole burning at low modulation frequency, and by Bragg-reflection from atomic population gratings at high frequency.

As our main interest is in laser stabilization, we should assess whether these signals are sufficient for our experiments.  As discussed above, a larger slope increases the overall servo gain, which reduces the amplitude of frequency fluctuations when locked. To set the relevant scale, we compare our results to the tuning transfer functions of our laser: $260\pm13$~MHz/V by piezo and $5000\pm250$~MHz/V by injection current.\footnote{Our diode tunes $250\pm12$~MHz/mA, and our current controller (Thorlabs LDC202C) has a modulation input with $1/(50\Omega)$ transconductance.}  Conservatively taking the piezo value, this sets the overall gain (without PID loop) to be $\sim 25$ for direct modulation and $\sim 150$ for modulation transfer.  With added gain from a PID controller, which can easily exceed 100 over our 200~kHz bandwidth, we therefore expect either method to be sufficient.  We should also consider the noise level, which was typically $\sim 15$~mV$_{\rm rms}$ for both methods under optimum conditions.  In terms of signal-to-noise ratio, both methods again seem sufficient.  However, taking the ratio of noise to slope gives estimated lower bounds on the short-term frequency fluctuations of $\sim 150$~kHz and $\sim 25$~kHz for direct modulation and modulation transfer, respectively.  Since these estimates neglect limitations due to finite servo bandwidth and additional technical noise, modulation transfer may be preferred for more demanding applications.  Finally, we consider the capture range, defined as the frequency range around the zero crossing over which the slope preserves its sign.  This was typically about 15~MHz for direct modulation, and 5~MHz for modulation transfer. The optimum capture range represents some compromise between tightness of lock and stability against perturbations. A preference for large or small capture range therefore depends on the details of the experimental arrangement and application.

In comparing the two methods it is clear that modulation transfer produces a far greater slope, but is unsuitable for locking in the vicinity of the crossover (B) feature.  Due to the strong suppression of the Doppler background, modulation transfer is expected to give a more well-defined lock point, in the sense of being very near the $F=2\to F^\prime=3$ resonance.  In contrast, the direct modulation signal exhibits contributions from numerous transitions, and the Doppler background causes an additional shift in the actual lock point.  However, as stated above, the small capture range of the modulation transfer signal, which is in many ways a virtue, may cause the laser to become more easily unlocked under the influence of environmental disturbances.  We therefore expect direct modulation to be more robust under such circumstances.

There are a number of further investigations which could be undertaken.  For example, it may be worth iteratively optimizing the vapour temperature and pump and probe intensities.  This could be especially useful for modulation transfer, which bears relatively little resemblance to saturated absorption.  We were also unable to perform a satisfactory optimization of the modulation frequency and sideband fraction, which can strongly affect the signal shapes and amplitudes \cite{Bjo83,McC08}.  For this purpose, it would be better to have a wideband EOM and high-voltage amplifier, which were not available to us. One can in principle also improve the signal by expanding the beams with a telescope \cite{McC08}.  In cases when one is limited by the beam intensity in the vapour, but not optical power at the detector, this can increase the signal amplitude and slope.

\ack
This work was funded by an EPSRC Science and Innovation award (EP/E036473/1).  We acknowledge useful discussions with S.~Cornish and members of the Cold Atoms group at the University of Birmingham, and thank P.~Petrov for providing feedback on the manuscript.  S.~Doravari advised on the laser design, and parts were built with expert technical assistance from J.~Dyne of the Centre for Cold Matter at Imperial College, London.  The EOM system was designed by J.~Kronjaeger and built with the assistance of N.~Meyer, and R.~Culver assisted with the Fabry-Perot and fibre ring cavities.

\end{document}